\documentclass{ws-ijmpb}

\usepackage[utf8]{inputenc}
\usepackage[english]{babel}
\usepackage{graphicx}
\usepackage{amsmath}
\usepackage{amssymb}
\usepackage{bbold}
\usepackage[abs]{overpic}

\newcommand{\eS}{\mathcal{S}}
\newcommand{\trace}{\text{Tr}}
\newcommand{\Id}{{\mathbb 1}}
\newcommand{\cibb}{{\mathbb c}}
\newcommand{\ket}[1]{\left\vert#1\right\rangle}

\begin{document}

\markboth{P. Silvi, D. Rossini, G.E. Santoro, R. Fazio, and V. Giovannetti}
{MPS representation for Slater determinants and CI states}

\title{Matrix Product State representation for Slater Determinants \\ and Configuration Interaction States}

\author{PIETRO SILVI}
\address{Institut f\"ur Quanteninformationsverarbeitung, Universit\"at Ulm, D-89069 Ulm, Germany\\
International School for Advanced Studies (SISSA), Via Bonomea 265, I-34136 Trieste, Italy}

\author{DAVIDE ROSSINI}
\address{NEST, Scuola Normale Superiore and Istituto di Nanoscienze - CNR,  Pisa, Italy}

\author{ROSARIO FAZIO}
\address{NEST, Scuola Normale Superiore and Istituto di Nanoscienze - CNR,  Pisa, Italy\\
Center for Quantum Technologies, National University of Singapore, Republic of Singapore}

\author{GIUSEPPE E. SANTORO}
\address{International School for Advanced Studies (SISSA), Via Bonomea 265, I-34136 Trieste, Italy\\
International Centre for Theoretical Physics (ICTP), P.O. Box 586, I-34014 Trieste, Italy\\
CNR-IOM Democritos National Simulation Center, Via Bonomea 265, I-34136 Trieste, Italy}

\author{VITTORIO GIOVANNETTI}
\address{NEST, Scuola Normale Superiore and Istituto di Nanoscienze - CNR,  Pisa, Italy}

\maketitle

\begin{abstract}

Slater determinants are product states of filled quantum fermionic orbitals.
When they are expressed in a configuration space basis chosen a priori, their entanglement is bound and controlled.
This suggests that an exact representation of Slater determinants as finitely-correlated states is possible. 
In this paper we analyze this issue and provide an exact Matrix Product representation for Slater determinant states. 
We also argue possible meaningful extensions that embed more complex configuration interaction states into the description.

\end{abstract}

%\pacs{03.67.-a, 05.30.-d, 71.15.-m}

% Quantum information, 03.67.-a
% Fermi-Dirac statistics, 05.30.-d
% Condensed matter, calculation methods, 71.15.-m


\section{Introduction}\label{sec:intro}

Starting from its first formulation in 1992, the density matrix renormalization group (DMRG) 
has rapidly established itself as the leading numerical method in the simulation of statical and 
dynamical properties of strongly correlated one-dimensional quantum systems\cite{White92,SchollwockDMRG}.
The versatility of this approach allowed the extension of DMRG applications to fields quite far from 
those it was originally designed to address, e.g. the quantum chemistry of small to medium-sized molecules\cite{Chem}.
A deeper understanding of the mechanisms which contribute in making DMRG such a successful tool 
has been achieved recently, when this technique was shown to be equivalent to a variational approach 
over the set of Matrix Product States (MPSs)\cite{PrimoMPS}.
The latter were originally named ``finitely correlated states'', due to their intrinsic ability in capturing 
the internal structure of many-body quantum states having exponentially decaying correlations\cite{FannesNacht}.
Remarkably it has been found that the ground state of the one-dimensional AKLT model 
is exactly given by a MPS\cite{AKLT,FannesNacht1}, thus stimulating a plethora of investigations 
in the quantum many-body realm 
(e.g. see Refs.~\refcite{Arealaws,MPSVidal,MPSground,QptMps,Schuck2008,MurgRev2008,CiracRev2009,SchollwockMPS}
and references therein).
The MPS architecture is so immediate and powerful that in some cases it is even amenable 
to analytical studies, providing an exact, and often \emph{optimal} representation for several classes 
of quantum states beyond the AKLT model, including GHZ, W, Majumdar-Gosh, Cluster\cite{MPSreps}, and Laughlin states\cite{IblisLaugh}.

In this paper we discuss how Slater determinants can be represented in terms of MPSs. 
Slater determinants correspond to  product states of (say) $N$ fermions where each fermion occupies 
a different single-particle orbital out of an orthonormal set (typically solution of a mean-field Hamiltonian). 
They are the starting point of most many-body calculations (e.g. {\it ab-initio} Hartree-Fock 
and post-Hartree-Fock methods)\cite{QChem} and, even though lacking true electron-electron correlations, 
their characterization poses significant problems when working on a configuration space 
for which the single-particle orbitals entering the determinant are not local (by this we mean that
the latter might have extended and non-mutually excluding supports in a wavefunction basis chosen a priori).
For this reason, most previous attempts to address quantum chemistry problems with DMRG\cite{QChem,DMRGAttempt}, 
and more in general with Tensor Network variational structures\cite{TNAttempt,Cayley}, 
agreed on adopting the Hartree-Fock basis as the selected configuration space for the correlation calculus. 
In the present work, we step back from this assumption, and show that that the description 
can be kept simple and manageable even when the chosen configuration basis is not related 
to the solutions of the one-body problem.

The paper is organized as follows: in Sec.~\ref{Sec:notations} we review the basic concepts
of fermionic quantum states and set the notations.
In Sec.~\ref{Sec:slater} we provide our prescription to describe a generic Slater
determinant state by means of a MPS. To this aim we first introduce a representation
of a single fermionic operator in terms of a Matrix Product Operator (MPO) %(Subsec.~\ref{Sec:MPOfermi}), 
and then exploit that design in order to build the Slater determinant MPS (Subsec.~\ref{Sec:MPOstack}).
In this framework it will then be easy to introduce a compact and numerically manageable representation 
for the many-body transformation which corresponds to the one-body basis change (Subsec.~\ref{Sec:Grid}).
In Sec.~\ref{Sec:optimal} we will prove the optimality of our description in a general setting,
through some considerations based upon quantum entanglement estimators.
A straightforward generalization to other particle-exchange statistics is given 
in Sec.~\ref{Sec:statistics}, while a possible extension to Configuration Interaction 
is described in Sec.~\ref{Sec:CI}.
Finally in Sec.~\ref{Sec:concl} we draw our conclusions.

\section{Basic notations}  \label{Sec:notations}

Consider a system composed of $N$ spinless fermions, which can occupy $L$ states (sites), identified by a canonical 
basis of one-body  wavefunctions, in some effective configuration (or momentum) space\cite{NOTA1}. 
As usual we can  account for the Fermi statistics of the system 
by  defining a complete ordering of the canonical basis and enforcing a Jordan Wigner Transformation
(JWT)\cite{JWT,Lieb1961,Shastry1986,Bravyi2002}, which is common ground when addressing
fermionic problem with numerical renormalization strategies\cite{VersCirJW}.
Accordingly, each lattice site, say the one labeled with the index $\ell$, is fully characterized 
by assigning a set of  Pauli 
matrices $\sigma^\alpha_\ell$ ($\alpha = x,y,z$), $\sigma_\ell^\pm = (\sigma_\ell^x \pm i \sigma_\ell^y)/2$,
and indicating with $\ket{0}_\ell$, $\ket{1}_\ell$ its empty/occupied level, which satisfy the relations
$\sigma_\ell^{z} \ket{0}_\ell = \ket{0}_\ell$,  $\sigma_\ell^{z} \ket{1}_\ell = -\ket{1}_\ell$ respectively. 
Then, following the standard procedure, we construct the correspondence:
\begin{eqnarray} \label{eq:Wigner1}
  \ket{\Omega} & \longrightarrow &  \ket{0}_1 \ket{0}_2 \ldots \ket{0}_L = \ket{00 \ldots 0} \, ,\\
  c_\ell        & \longrightarrow & \cibb_\ell \equiv \sigma^{z}_{1} \otimes \ldots 
  \otimes \sigma^{z}_{\ell-1} \otimes \sigma^{+}_\ell \otimes \Id_{\ell+1} \otimes \ldots \otimes \Id_L \, ,
  \nonumber \\
  c_\ell^{\dagger} c_\ell & \longrightarrow & \cibb_{\ell}^{\dagger} \cibb_{\ell} = (1 - \sigma^{z}_\ell)/2 \, ,
  \nonumber
\end{eqnarray}
where $\ket{\Omega}$ is the vacuum (Fock) state of the fermionic system, 
while $c^\dagger_\ell$ ($c_\ell$) is the creation (destruction) operator which creates (annihilates) 
a fermion on the $\ell$-th one-body state of the canonical basis, and obeys 
the canonical anti-commutation relations
$\{c_\ell^\dagger, c^\dagger_m\} =\{c_\ell, c_m\} = 0$, $\{c_\ell, c^{\dagger}_m\} = \delta_{\ell,m}$.
As long as the set of $L$ one-body levels is complete, any Fermionic state
$|\Psi\rangle$ can now be expressed as a many-body state $|\Psi^{(JW)} \rangle$ of  $L$ qubits via the following mapping:
\begin{multline} \label{eq:cantens}
  \ket{\Psi} = \sum_{s_1 \in \{0,1\}} \ldots \sum_{s_L \in \{0,1\}} \mathcal{T}_{s_1 \ldots s_L} \; 
  (c_{1}^{\dagger})^{s_1} \ldots (c_{L}^{\dagger})^{s_L} \ket{\Omega}
  \\
  \longrightarrow \; \vert \Psi^{(JW)} \rangle
  \equiv \sum_{\{s_j\} \in \{0,1\}}  \mathcal{T}_{s_1 \ldots s_L} \; \ket{s_1 \ldots s_L},
\end{multline}
where, in the first line,  $(c_{\ell}^{\dagger})^{0}$ stands for the identity operator $\Id$, while 
the creation operators $c_{\ell}^{\dagger}$ are placed according to the chosen ordering.
The tensor  $\mathcal{T}_{s_1 \ldots s_L}$  holds the quantum amplitudes of every fermionic occupancy configuration.

A Slater determinant state $|\Sigma\rangle$ is a special $N$-particle state 
that can be cast in the form 
\begin{equation} \label{eq:Slaz}
  |\Sigma\rangle = \tilde{c}_1^{\dagger}\,\tilde{c}_2^{\dagger}\,\ldots \tilde{c}_{N}^{\dagger}\,| \Omega \rangle,
\end{equation}
where the $\tilde{c}_{\alpha}^{\dagger}$ are creation operators associated to a collection 
of mutually orthogonal one-body orbitals. These
are eventually delocalized with respect to the original canonical basis.
In particular, we decompose the $\tilde{c}_{\alpha}^{\dagger}$s according to
\begin{equation} \label{defctilde}
  \tilde{c}_{\alpha}^{\dagger} = \sum_{\ell = 1}^{L} \phi_{\alpha}(\ell)\; c^{\dagger}_{\ell}\;,
\end{equation}
with $\phi_{\alpha}(\ell)$ satisfying the orthonormality relations
\begin{eqnarray}
  \sum_{\ell=1}^L \phi_{\alpha}(\ell) \phi^{\star}_{\beta}(\ell) = \delta_{\alpha, \beta}\;.
\end{eqnarray} 
The explicit determinant form of the many-particle state $|\Sigma\rangle$ becomes manifest when expanding the 
product $\tilde{c}_1^{\dagger}\,\tilde{c}_2^{\dagger}\,\ldots \tilde{c}_{N}^{\dagger}$ in terms of the $c_\ell^\dag$s. 
Indeed simple mathematical manipulations allow one to write 
\begin{equation} \label{eq:slasla}
  |\Sigma\rangle = \!\!\!\! \sum_{1 \leq \ell_1 < \ldots < \ell_{N} \leq L}
  \left| \begin{array}{ccc}
    \phi_{1}(\ell_1) & \cdots & \phi_{1}(\ell_N) \\
    \vdots  & \ddots & \vdots \\
    \phi_{N}(\ell_1) & \cdots & \phi_{N}(\ell_N) \\
  \end{array} \right|
  c_{\ell_{1}}^{\dagger} \ldots c_{\ell_{N}}^{\dagger} |\Omega\rangle \,,
\end{equation}
which can be cast in the form of Eq.~\eqref{eq:cantens} by taking
\begin{eqnarray}\label{fdNUOVA}
  \mathcal{T}_{s_1 \ldots s_L} = 
  \!\!\!\! \sum_{1 \leq \ell_1 < \ldots < \ell_{N} \leq L}
    \left| \begin{array}{ccc}
    \phi_{1}(\ell_1) & \cdots & \phi_{1}(\ell_N) \\
    \vdots  & \ddots & \vdots \\
    \phi_{N}(\ell_1) & \cdots & \phi_{N}(\ell_N) \\
  \end{array} \right|\;
  \prod_{\ell' \in \{\ell_q \}_q} \!\!\! \delta_{s_{\ell'}, 1}
  \prod_{\ell'' \in \{\ell_q \}_q^{\text{c}} } \!\!\! \delta_{s_{\ell''}, 0}\;,
\end{eqnarray} 
where $\ell'$ spans over the site indices which appear in the determinant, and $\ell''$ on those not appearing.
Equation~\eqref{fdNUOVA} explicitly manifests the complexity of representing $\ket{\Sigma}$ in
qubit-oriented configuration space.

%%%%%%%%%%%%%%%%%%%%%%%%%%%%%%%%%%%%%%%%%%%%%%%%%%%%%%%%%%%%%%%%%
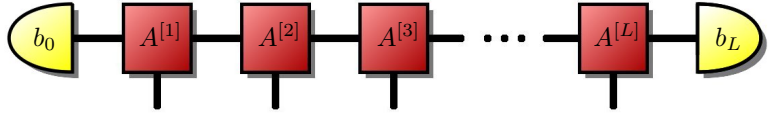
\begin{figure}[t]
  \begin{center}
  \begin{overpic}[width = 300pt, unit=1pt]{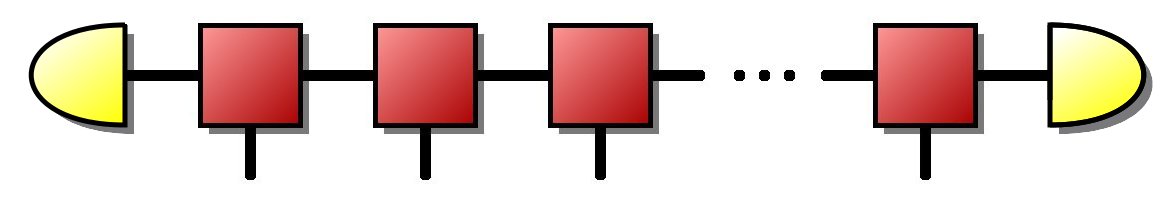}
    \put(17, 29){$b_0$}
    \put(57,29){$A^{[1]}$}
    \put(102, 29){$A^{[2]}$}
    \put(147, 29){$A^{[3]}$}
    \put(229, 29){$A^{[L]}$}
    \put(275, 29){$b_L$}
  \end{overpic}
  \end{center}
  \caption{ \label{fig:boumps}
    (Color online). MPS representation of the tensor $\mathcal{T}_{s_1 \ldots s_L}$ appearing 
    in Eq.~\eqref{eq:cantens}. Every element $A^{[\ell]}$ is a three-index complex tensor, 
    graphically depicted as a red box. The two yellow elements represent the correlation boundary vectors.
    Open links, the vertical ones, refer to physical degrees of freedom (sites); while
    the closed, horizontal links, are the correlation space indices, which are contracted through
    the matrix product.
    }
\end{figure}
%%%%%%%%%%%%%%%%%%%%%%%%%%%%%%%%%%%%%%%%%%%%%%%%%%%%%%%%%%%%%%%%%

The purpose of the paper is to provide an exact and efficient representation for states of the 
form~\eqref{eq:slasla}, \eqref{fdNUOVA} in the MPS formalism.
In practice this consists in writing 
$\mathcal{T}_{s_1 \ldots s_L}$ as a contraction of smaller elements according to the decomposition
\begin{equation} \label{eq:MPSampli}
  \mathcal{T}_{s_1 \ldots s_L} = 
  (b_0| A_{s_{1}}^{[1]} \cdot A_{s_{2}}^{[2]} \cdot \ldots \cdot A_{s_{L}}^{[L]} |b_L) \, ,
\end{equation} 
where $A_{s}^{[\ell]}$ are a set of $D \times D$ matrices, while $|b_L)$ and $(b_0|$ are respectively
a $D$-dimensioned column vector, and a $D$-dimensioned linear functional (row vector)\cite{NOTANEW1}.
A pictorial representation of this identity, which follows the standard graphical convention\cite{MurgRev2008},
is presented in Fig.~\ref{fig:boumps}.
In Eq.~(\ref{eq:MPSampli}) $D$ is a free parameter quantifying the degree of refinement 
of the MPS representation (thus also referred to as \emph{refinement parameter}, 
as well as \emph{bond dimension}). 
As detailed in Sec.~\ref{Sec:optimal}, we will focus on an explicit construction~(\ref{eq:MPSampli}) 
which ensures the \emph{smallest} possible value of $D$, regardless of the shape 
of the orbitals $\phi_{\alpha}(\ell)$ entering in Eq.~(\ref{fdNUOVA}).
We will show that this minimal (or optimal) bond dimension is $D = 2^{N}$.

\section{MPS representation of a Slater determinant}  \label{Sec:slater} 

\subsection{MPOs for delocalized Fermi operators}     \label{Sec:MPOfermi}

\begin{figure}[b]
  \begin{center}
  \begin{overpic}[width = 300pt, unit=1pt]{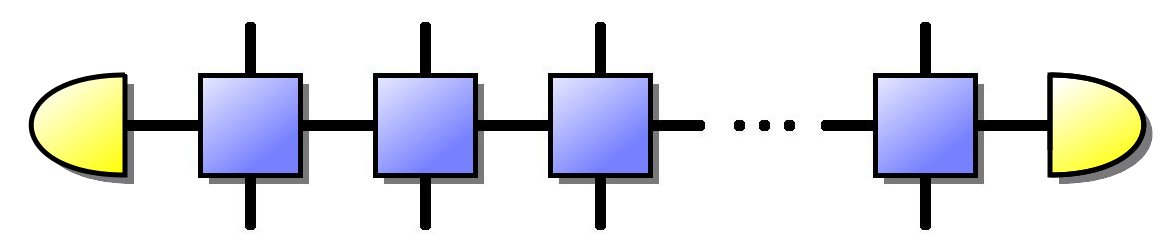}
    \put(17, 29){$\beta_0$}
    \put(57, 29){$B^{[1]}$}
    \put(102, 29){$B^{[2]}$}
    \put(147, 29){$B^{[3]}$}
    \put(229, 29){$B^{[L]}$}
    \put(274, 29){$\beta_L$}
  \end{overpic}
  \end{center}
  \caption{ \label{fig:boumpo}
    (Color online). MPO representation of the fermionic operator $\tilde{c}_\alpha^\dag$, as
    appearing in Eq.~\eqref{eq:CreatMPO}. Every blue box $B^{[\ell]}$ is a four-index complex tensor.}
\end{figure}
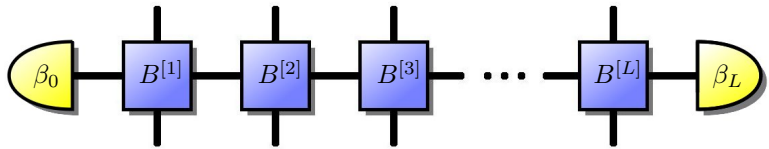

To find a MPS representation~(\ref{eq:MPSampli})  for $|\Sigma\rangle$ it is instrumental to first provide a MPO 
description\cite{MPOMurg} for the Fermi operators introduced in Eq.~\eqref{defctilde}.
According to this goal, for each lattice site $\ell$ and for each orbital $\alpha$, we identify a collection of four 
$D\times D$ matrices ${B^{[\ell]}}_{0}^{0}, {B^{[\ell]}}_{0}^{1}, {B^{[\ell]}}_{1}^{0}, {B^{[\ell]}}_{1}^{1}$ 
and boundary $D$-dimensional vectors $( \beta_0 |$ and $| \beta_L)$, 
such that the creation operators $\tilde{c}_{\alpha}^{\dagger}$ can be decomposed as
\begin{equation} \label{eq:CreatMPO}
  \tilde{c}_{\alpha}^{\dagger} = \!\!\! \sum_{s_1 \ldots s_L = 0}^1 \; \sum_{r_1 \ldots r_L = 0}^1
  ( \beta_0 | \vphantom{\sum} {B^{[1]}}_{s_1}^{r_1} \cdot \ldots \cdot {B^{[L]}}_{s_L}^{r_L} | \beta_L) \;\;
  (c_{1}^{\dagger})^{s_1} \ldots (c_{L}^{\dagger})^{s_L} |\Omega\rangle \langle \Omega |
  \,c_{L}^{r_L} \ldots c_{1}^{r_1}.
\end{equation}
The tensor 
\begin{eqnarray} \label{tensorEMME}
{\cal M}_{s_1,\cdots,s_L}^{r_1,\cdots, r_L} := 
( \beta_0 | \vphantom{\sum} {B^{[1]}}_{s_1}^{r_1} \cdot \ldots \cdot {B^{[L]}}_{s_L}^{r_L} | \beta_L)\;,
\end{eqnarray} 
entering the above expression is the MPO representation of $\tilde{c}_{\alpha}^{\dagger}$
depicted in Fig.~\ref{fig:boumpo}.
%
%%%%%%%%%%%%%%%%%%%%%%%%%%%%%%%%%%%%%%%%%%%%%%%%%%%%%%%%%%%%%%%%%
%%%%%%%%%%%%%%%%%%%%%%%%%%%%%%%%%%%%%%%%%%%%%%%%%%%%%%%%%%%%%%%%%
%
At the level of the JW mapping~\eqref{eq:cantens} this correspond to 
decompose the anticommuting qubit-orbital construction operators in the following formalism:
\begin{equation} \label{eq:CreatMPO2}
  \tilde{\mathbb c}_\alpha^\dag = \sum_{s_1 \ldots s_L = 0}^1 \; \sum_{r_1 \ldots r_L = 0}^1
  ( \beta_0 | \vphantom{\sum} {B^{[1]}}_{s_1}^{r_1} \cdot \ldots \cdot {B^{[L]}}_{s_L}^{r_L} | \beta_L) \;,
  |s_1 \ldots s_L\rangle\langle r_1 \ldots r_L| \;,
\end{equation}
where such operators are defined by
\begin{equation} \label{eq:firstreux}
  \tilde{\cibb}_{\alpha}^{\dagger}
  = \sum_{\ell} \phi_{\alpha}(\ell)\, \cibb_{\ell}^{\dagger} =
  \sum_{\ell} \phi_{\alpha}(\ell) \left[ \sigma_{1}^{z} \otimes \ldots \otimes \sigma_{\ell-1}^{z}
    \otimes \sigma_{\ell}^{-} \otimes \Id_{\ell+1} \otimes \cdots \otimes  \Id_{L} \right].
\end{equation}
Elaborating from the results of Ref.~\refcite{MPOFrowis},
we claim that to build such representation it is sufficient to take $D = 2$. Indeed, assume to add a two-level
fictitious degree of freedom, with canonical basis states $|0)$ and $|1)$. Then it is possible to write the previous
expression \eqref{eq:firstreux} as a contraction over this new degree of freedom of a matrix product:
\begin{equation} \label{eq:firstreux2}
  \tilde{\cibb}_{\alpha}^{\dagger} =
  (1| \prod_{\ell = 1 \leftarrow}^{\rightarrow L} \left[ \vphantom{\sum}
    \Id_\ell \otimes |0)(0| +
    \sigma^{z}_\ell \otimes |1)(1| +
    \phi_{\alpha}(\ell) \sigma_{\ell}^{-} \otimes |1)(0|
    \right] |0)
  = (1| {\prod_{\ell = 1 \leftarrow}^{\rightarrow L} \mathcal{B}_{\ell}} |0)
\end{equation}
where the product is sorted in increasing order in $\ell$ from left to right. 
The $\mathcal{B}_{\ell}$ terms in Eq.\eqref{eq:firstreux2} are meant as matrices 
in both the $\ell$-th site and the common fictitious degrees of freedom, and read
\begin{equation}
  \mathcal{B}_{\ell} = \frac{1}{2} \left( \Id_{\ell} + \sigma^{z}_{\ell} \right) \otimes
  \left( \Id_{f} + \sigma^{z}_{f} \right) - \sigma^{z}_{\ell} \otimes \sigma^{z}_{f} +
  \phi_{\alpha}(\ell) \,\sigma_{\ell}^{-} \otimes \sigma_{f}^{-}.
\end{equation}
To recover the formalism in \eqref{tensorEMME}, we just need to write $\mathcal{B}_{\ell}$ 
entrywise in the real degree of freedom. When doing this we obtain
\begin{equation} \label{eq:FermiMPOrecipe}
  \begin{array}{cc}
    {B^{[\ell]}}_{0}^{0} = \Id_f
    & {B^{[\ell]}}_{0}^{1} = 0
    \vspace{1.5mm} \\
       {B^{[\ell]}}_{1}^{0} = \phi_{\alpha}(\ell) \, \sigma^{-}_f
       &   {B^{[\ell]}}_{1}^{1} = \sigma^{z}_f
    \vspace{1.5mm} \\
       | \beta_L ) = \left( \begin{array}{cc}
         1 \\ 0
       \end{array} \right) = |0)
       & \quad ( \beta_0 | = \left( \vphantom{{B^{[\ell]}}_{1}} 0\;\;\;1 \right) = (1|.
  \end{array}
\end{equation}
%1
It is easy to see that the information on the orbital
$\alpha$, via the orbital wave-function $\phi_{\alpha}(\ell)$,
enters only in one of the 16 elements of the four-index tensor $\mathcal{B}_{\ell}$: apart from that, the
expression~\eqref{eq:FermiMPOrecipe} is formally homogeneous in $\ell$ and no dependence on $\alpha$ is
left on the boundary elements $( \beta_0 |$ and $| \beta_L)$.

In a similar way we can also construct the MPO representation for the annihilation operators $\tilde{c}_{\ell}$. 
Alternatively we find it useful to recall that the MPO representation behaves very well under
the adjoint application: in practice, if the matrices ${B^{[\ell]}}^{r}_{s}$ provides a MPO representation 
for the operator $\Theta$ then ${B^{*[\ell]}}^{s}_{r}$ do the same for the adjoint operator $\Theta^{\dagger}$ 
(where $x^*$ denotes complex conjugation of $x$),
i.e.
\begin{equation} \label{eq:AnnihMPO}
  \tilde{c}_{\alpha} = \!\!\! \sum_{s_1 \ldots s_L = 0}^1 \; \sum_{r_1 \ldots r_L = 0}^1
  ( \beta_0 | \vphantom{\sum} {B^{* [1]}}_{r_1}^{s_1} \cdot \ldots \cdot {B^{* [L]}}_{r_L}^{s_L} | \beta_L)  \;\;
  (c_{1}^{\dagger})^{s_1} \ldots (c_{L}^{\dagger})^{s_L} |\Omega\rangle \langle \Omega |
  \,c_{L}^{r_L} \ldots c_{1}^{r_1}.
\end{equation}

As an final remark, we notice that, while still keeping the matrices $B^{[\ell]}$ entering~\eqref{tensorEMME} 
as in Eq.~\eqref{eq:FermiMPOrecipe} it is possible to modify the action of the associated operator by simply  
changing the boundary vectors $(\beta_0|$ and $(\beta_L|$. For instance, by identifying 
both $|\beta_0)$ and $|\beta_L)$ with $|0)$, the resulting MPO will represent the identity operator. 
More generally, manipulating boundary vectors gives
\begin{equation}
  \begin{array}{cc|ccc}
    (\beta_0| & |\beta_L) & \quad \mbox{operator (spin)} & \mbox{(Fermi)} & \mbox{(name)}
    \\ \hline
    (1| & |0) & \tilde{\cibb}^{\dagger}_{\alpha} & \tilde{c}^{\dagger}_{\alpha} & \mbox{Construction} \\    
    (0| & |0) & \Id^{\otimes L} & \Id^{\otimes L} & \mbox{Identity} \\
    (1| & |1) & ({\sigma^{z}})^{\otimes L} & (\Id - 2 \,c^{\dagger} c)^{\otimes L} & \mbox{Parity} \\
    (0| & |1) & 0 & 0 & \mbox{Null.}
  \end{array}
  \label{eq:MPObound}
\end{equation}
In particular, when $|\beta_L) = |0)$ one can then regard the left correlation space boundary 
as a local switch that ``activates'' or ``deactivates'' the creation operator.

%%%%%%%%%%%%%%%%%%%%%%%%%%%%%%%%%%%%%%%%%%%%%%%%%%%%%%%%%%%%%%%%%
\begin{figure}[t]
  \begin{overpic}[width = \columnwidth, unit=1pt]{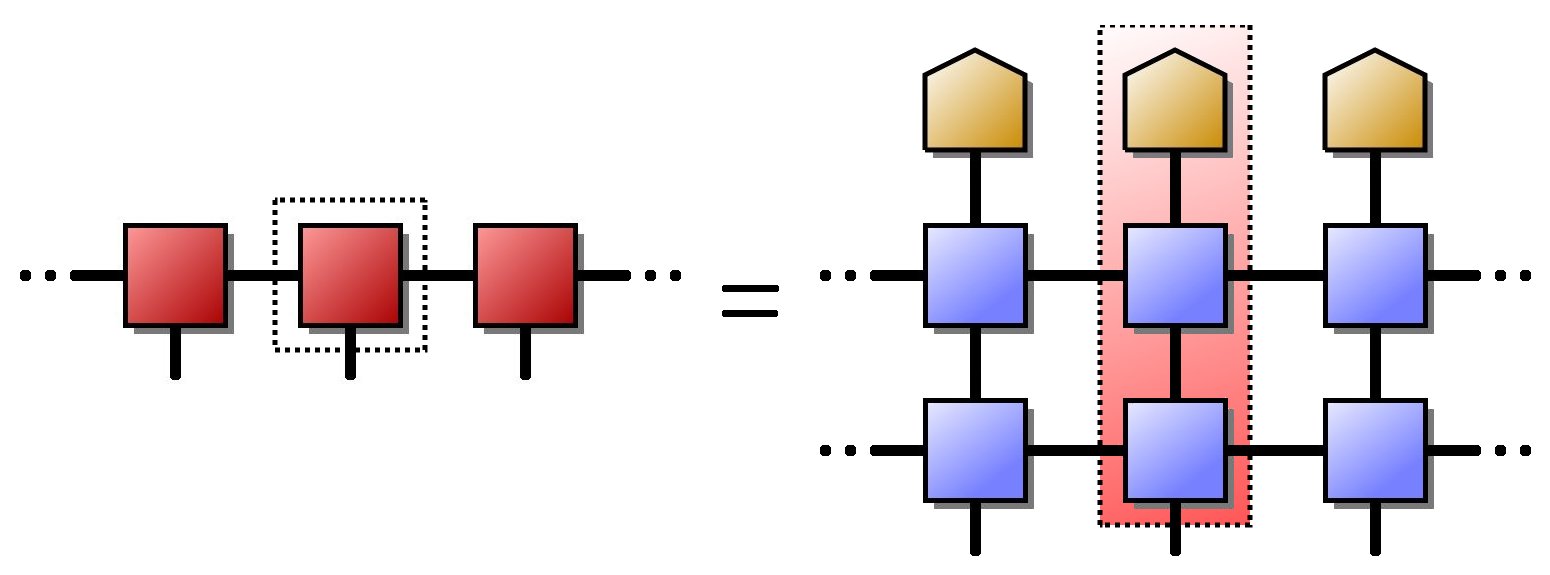}
    \put(75, 67){${A^{[\ell]}}$}
    \put(267, 67){${B^{[\ell]}}$}
    \put(271, 105){$0$}
  \end{overpic}
  \caption{ \label{fig:stacker}
    (Color online). The MPS construction provided in Eq.~\eqref{eq:Slamprecipe2} is obtained
    by vertically stacking Fermi operators in their MPO form. Thus an $A^{[\ell]}$ tensor
    is the contracted column of $B^{[\ell]}$ tensors applied to the local vacuum state $|0\rangle$.
  }
\end{figure}
%%%%%%%%%%%%%%%%%%%%%%%%%%%%%%%%%%%%%%%%%%%%%%%%%%%%%%%%%%%%%%%%%

\subsection{Stacking Fermi operators into a MPS}  \label{Sec:MPOstack}

Let us now focus on the Slater determinant state $\ket{\Sigma}$ in Eq.~\eqref{eq:Slaz},
following a prescription analogous to the one introduced for MPS representations
of algebraic Bethe ansatz\cite{Algebethe}.
The vacuum state $|\Omega\rangle \rightarrow |0\ldots0\rangle$ is a MPS with $D = 1$,
since it is a product state.
It is clear that, when we apply $\tilde{c}_{\alpha}^{\dagger}$ to a given MPS $|\Phi_{\text{MPS}}\rangle$,
we can exploit its MPO formalism presented in the previous subsection, and immediately obtain 
a MPS expression for $|\Phi' \rangle = \tilde{c}_{\alpha}^{\dagger} |\Phi_{\text{MPS}}\rangle$.
A sketch of the procedure is depicted in Fig.~\ref{fig:stacker}

This process can be repeated recursively, every time adding a new $\tilde{c}_{\alpha}^{\dagger}$
until the whole operators string $\tilde{c}_{1}^{\dagger} \ldots \tilde{c}_{N}^{\dagger}$ is included, 
and acting on $|\Omega\rangle$.
This is equivalent to writing the following three-index tensor, which represents
the stacking of $B^{[\ell]}$ tensors applied to the local vacuum coming from the Fock state $\ket{\Omega}$:
\begin{equation} \label{eq:Slamprecipe}
  A^{[\ell]}_s = \sum_{q_2 \ldots q_N = 0}^{1}
  \left(  {B^{[\ell,1]}}_{s}^{q_2} \otimes {B^{[\ell,2]}}_{q_2}^{q_3}
  \otimes \ldots \otimes {B^{[\ell,N]}}_{q_N}^{0}
  \right) \; ,
\end{equation}
here the ${B^{[\ell,\alpha]}}_{s}^{r}$ compose the set of $2 \times 2$ matrices introduced 
in Eqs.~\eqref{eq:FermiMPOrecipe}, concerning the orbital $\phi_\alpha$.
This description uses as a whole a total refinement parameter (also called bond-link dimension)
of $D = 2^{N}$; exponential in the fermion number, but disregarding the total size $L$.
By substituting~\eqref{eq:FermiMPOrecipe} into Eq.~\eqref{eq:Slamprecipe}, and exploiting 
the fact that $B_{0}^{1}$ is always the null matrix, the whole equation is reduced to
\begin{equation} \label{eq:Slamprecipe2}
  \begin{aligned}
    A^{[\ell]}_0 &= \Id_{2^N \times 2^N}
    \vphantom{\int} \\
    A^{[\ell]}_1 &= \sum_{\alpha = 1}^{N} \phi_{\alpha}(\ell)
    \left( {\sigma^z}^{\otimes {\alpha -1}} \otimes \sigma^{-} \otimes {\Id}_{2 \times 2}^{\otimes {N-\alpha}}
    \right).
  \end{aligned}
\end{equation}
The correlation boundary vectors are similarly constructed:
\begin{equation} \label{eq:Boundrecipe}
|b_{L}) = |\beta_L)^{\otimes N}  \quad {\rm and} \qquad (b_0| = (\beta_0|^{\otimes N} \, ,
\end{equation}
i.e. $|b_{L}) = |0)^{\otimes N} = \left( 1\:0\:0 \cdots 0 \right)^{\text T}$ and
$(b_0| = (1|^{\otimes N} = \left( 0 \cdots 0\:0\:1 \right)$.
With the prescriptions introduced in Eqs.~\eqref{eq:Slamprecipe2} and~\eqref{eq:Boundrecipe}
and a little algebraic manipulation, the reader can check that such MPS provides the correct amplitudes
for the Slater Determinant state
\begin{equation} \label{eq:Slamp}
  |\Sigma\rangle = 
  \sum_{s_1 \ldots s_L = 0}^1
  (b_0| A_{s_{1}}^{[1]} \cdot \ldots \cdot A_{s_{L}}^{[L]} |b_L)\;
  (c_{1}^{\dagger})^{s_1} \ldots (c_{L}^{\dagger})^{s_L} |\Omega\rangle.
\end{equation}

\vspace{.5em}
\emph{\textbf{Example --}} 
We would like to end this section by showing a simple example about this representation at work.
Let us consider the case of $N = 2$ fermions occupying two distinct orbitals.
In this scenario $A^{[\ell]}_{0}$ is the $4 \times 4$ identity operator, while
$A^{[\ell]}_{1} = \phi_1(\ell) [\sigma^{-} \otimes \Id] + \phi_2(\ell) [\sigma^z \otimes \sigma^- ]$, i.e.
\begin{equation}
  N = 2 \; \longrightarrow \;
  A^{[\ell]}_{1} = \left(
  \begin{array}{cccc}
    0 & 0 & 0 & 0 \\
    \phi_2(\ell) & 0 & 0 & 0 \\
    \phi_1(\ell) & 0 & 0 & 0 \\
    0 & \phi_1(\ell) & -\phi_2(\ell) & 0 \\
  \end{array} \right).
\end{equation}
The only products of matrices leading to nonzero amplitudes are those where exactly two
excitations $|1\rangle$ are present. Thus the sum in Eq.~\eqref{eq:Slamp} reduces to
\begin{eqnarray}
  \nonumber
    |\Sigma_2\rangle & = & \sum_{\ell_1 < \ell_2} \left( \phi_1(\ell_1)\, \phi_2(\ell_2) - \phi_1(\ell_2)\, \phi_2(\ell_1)
     \right) \, c_{\ell_1}^{\dagger} c_{\ell_2}^{\dagger} \,|\Omega\rangle\\
    & = & \sum_{\ell_1 < \ell_2}
    \left| \begin{array}{cc}
      \phi_1(\ell_1) & \phi_1(\ell_2) \\
      \phi_2(\ell_1) & \phi_2(\ell_2) \\
    \end{array} \right|
    \, c_{\ell_1}^{\dagger} c_{\ell_2}^{\dagger} \,|\Omega\rangle \, ,
\end{eqnarray}
where we have explicitly recovered the determinant expression, thus confirming
the validity of the MPS construction~\eqref{eq:Slamprecipe2} and~\eqref{eq:Boundrecipe}.

\subsection{Tensor network representation of one-body wavefunction basis change} \label{Sec:Grid}

When we derived the MPO representation~\eqref{eq:FermiMPOrecipe} for $\tilde{c}_{\alpha}^{\dagger}$, we also mentioned
that it is possible to control its overall action by adjusting one of the two correlation boundary vectors,
say the left one $(\beta_0|$ [see Eqs.~\eqref{eq:MPObound}].
In particular, the MPO produces $\tilde{c}_{\alpha}^{\dagger}$ if $(\beta_0| = (1|$, 
while it coincides with the identity operator $\Id$ as we set $(\beta_0| = (0|$.
Recall that the fermionic orbitals $\phi_{\alpha}(\ell)$ we adopted for the Slater determinant state 
formed an orthonormal set: let us complete it to an orthonormal basis $\{\phi_{\alpha}(\ell)\}_{\alpha}$, 
with $\alpha \in \{1 \ldots L\}$. The dimension must be $L$ by the assumption that the original set 
of $L$ wavefunctions was complete. As before, Eq.~\eqref{eq:FermiMPOrecipe} provides the construction 
operator $c_{\alpha}^{\dagger}$ for any of those orbitals $\phi_{\alpha}(\ell)$.

Now we stack together the MPOs, like we did for the Slater state, but instead of using only $N$ of them,
we stack the complete set, ordered from $\alpha = 1$ (on top) to $\alpha = L$ (at the bottom).
Moreover, instead of using the standard left correlation boundary vector $(\beta_0|_\alpha = (1|$ 
we set a generic $(\beta_0|_\alpha = (q_{\alpha}|$, where $q_{\alpha}$ are classical binary variables.
It is clear that the many-body operator $\Theta$ generated from this setup is equivalent to
\begin{equation}
  \Theta =
  \tilde{c}_{1}^{\dagger\, q_1} \; \tilde{c}_{2}^{\dagger\, q_2} \;
  \tilde{c}_{3}^{\dagger\, q_3} \ldots \tilde{c}_{L}^{\dagger\, q_L}.
\end{equation}
Finally, we apply such operator $\Theta$ to the vacuum $|\Omega\rangle$.
By doing so we are actually defining an application
from the binary strings of $\{q_{\alpha}\}_{\alpha}$ to the Fermi system:
\begin{equation} \label{eq:Wigner2}
  (q_1 \ldots q_L| \longrightarrow |\Psi_{\vec{q}} \rangle
  = \tilde{c}_{1}^{\dagger\, q_1} \; \tilde{c}_{2}^{\dagger\, q_2}
  \ldots \tilde{c}_{L}^{\dagger\, q_L} |\Omega\rangle.
\end{equation}
By linearity, this map extends to all the space generated by $(q_1 \ldots q_L|$, which corresponds to
the whole correlation bond-link space, $(q_1 \ldots q_L|$ being its canonical product basis. 
The map is clearly invertible, and its inverse is basically a JWT from the Fermi space
to the qubit lattice.
Only, this time the extended orbital set $\phi_\alpha$ has been chosen as basis of one-body wavefunctions,
so it is formally similar to~\eqref{eq:Wigner1}, but the one-body basis has \emph{changed}
(the old one is associated to the $c$, the new one to the $\tilde{c}$).

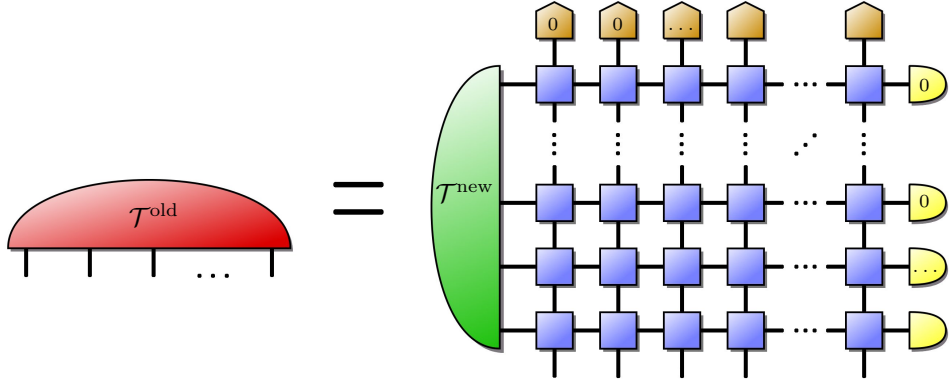
\begin{figure}[t]
  \begin{overpic}[width = \columnwidth, unit=1pt]{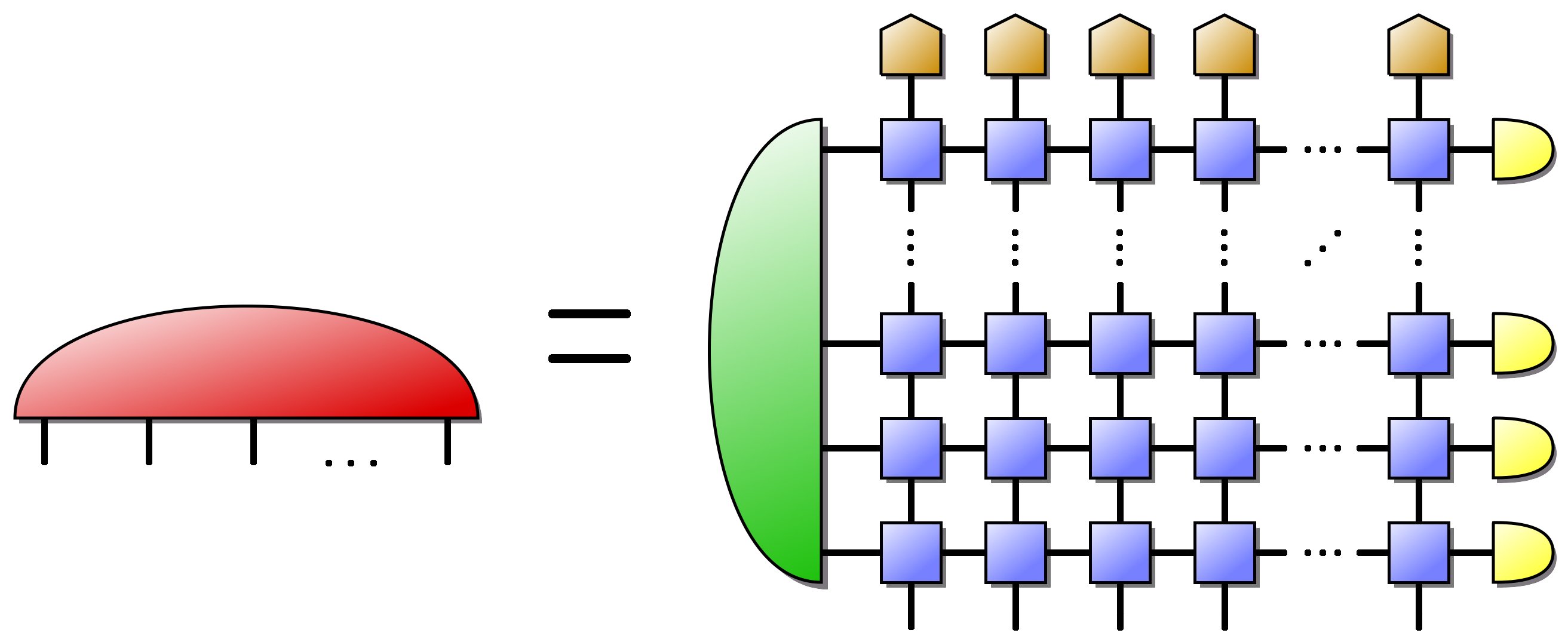}
    \put(49, 60){$\mathcal{T}^{\text{old}}$}
    \put(165, 69){$\mathcal{T}^{\text{new}}$}
    \put(207.5, 134) {\scriptsize $0$}
    \put(232, 134) {\scriptsize $0$}
    \put(253, 134) {\scriptsize $\ldots$}
    \put(348, 111) {\scriptsize $0$}
    \put(348, 67) {\scriptsize $0$}
    \put(345.5, 42.5) {\scriptsize $\ldots$}
  \end{overpic}
  \caption{ \label{fig:Grid}
    (Color online). Tensor network representation for the many-body state transformation of one-body basis change.
    $\mathcal{T}^{\text{old}}$ and $\mathcal{T}^{\text{new}}$ from Eq.~\eqref{eq:oldtonew} are tied
    through this pictorial equation. Blue boxes correspond to $B^{[\ell, \alpha]}$ tensors, where $\ell$ and $\alpha$
    respectively label the horizontal and vertical axis position in the grid, starting from the
    left-bottom corner.
  }
\end{figure}

In conclusion, we could use the extended MPO stack formalism to represent a transformation
of the many body state $|\Psi\rangle$
corresponding to a change of the chosen basis of one-body wavefunctions. I.e.~assuming that we can write
$|\Psi\rangle$ in both of the following expansions
\begin{equation} \label{eq:oldtonew}
    |\Psi\rangle = \!\!\! \sum_{s_1 \ldots s_L = 0}^{1} \mathcal{T}_{s_1 \ldots s_L}^{\text{old}}\,
    ({c}_{1}^{\dagger})^{s_1} \ldots ({c}_{L}^{\dagger})^{s_L} |\Omega\rangle = \!\!\!
    \sum_{q_1 \ldots q_L = 0}^{1} \mathcal{T}_{q_1 \ldots q_L}^{\text{new}}\,
    ({\tilde{c}}_{1}^{\dagger})^{q_1} \ldots ({\tilde{c}}_{L}^{\dagger})^{q_L} |\Omega\rangle,
\end{equation}
then the two components tensors $\mathcal{T}^{\text{old}}$ and $\mathcal{T}^{\text{new}}$ satisfy the
relation depicted in Fig.~\ref{fig:Grid}.

%%%%%%%%%%%%%%%%%%%%%%%%%%%%%%%%%%%%%%%%%%%%%%%%%%%%%%%%%%%%%%%%%
%%%%%%%%%%%%%%%%%%%%%%%%%%%%%%%%%%%%%%%%%%%%%%%%%%%%%%%%%%%%%%%%%

As a concluding remark, we recall that 2D tensor lattices cannot be efficiently contracted exactly in general.
Nevertheless, the tensor network that we built in Fig.~\eqref{fig:Grid} is made of sparse
tensors (the high amount of zero tensor elements due to particle-conservation symmetry relations), thus
effectively making it an easily contractible network.

\section{Entanglement and optimality of the description}  \label{Sec:optimal}

The mere existence of an exact finitely-correlated-state representation for a generic Slater Determinant $|\Sigma\rangle$,
automatically defines an \emph{upper bound} to the entanglement  contained in $|\Sigma\rangle$.
Indeed, MPS manifest a well-known bound\cite{MPSreps} to their entanglement entropy
(the von Neumann entropy of a subsystem-reduced density matrix $\rho$) given by
\begin{equation} \label{eq:entbound}
  \eS_{\text{VN}}(\rho) \equiv - \trace \left[ \rho \log \rho \right]
  \leq \log D.
\end{equation}
Clearly, any class of states allowing an exact MPS representation with established (eventually non-scaling)
bond dimension $D$, undergoes the same upper bound.

We now argue that, if no further information upon the filled orbitals $\phi_{\alpha}(\ell)$
within $|\Sigma\rangle$ is assumed, the exact representation we just gave,
Eqs.~\eqref{eq:Slamprecipe2} and~\eqref{eq:Boundrecipe}, is the most efficient in terms of MPS.
By this, we mean that there cannot be another general MPS representation $\tilde{A}$ for arbitrary $|\Sigma\rangle$
spending a bond-link dimension $D < 2^N$. To prove this, we will consider a specifically
built $N$-fermions Slater Determinant, and show that its entanglement entropy is equal to $\eS_{\text{VN}} = N \log 2$
(here we are assuming that $L \geq 2N$);
since any exact MPS representation for this state should have $D \geq 2^N$ due to~\eqref{eq:entbound},
we must conclude that $\tilde{A}$ fails to reproduce this state, and hence it is not general.
The present entanglement bound estimator can be perceived as a special case of
those treated in Refs.~\refcite{Ingo,IblisLaugh}, but it was developed autonomously.

For this setting, let us adopt a special set of ($L/2$-periodic) plane-waves
$\phi_{\alpha}(\ell) = L^{-1/2} \exp(4\pi i \alpha \ell / L)$. Now we consider a half-system partition, and
define a new double set of orbitals $\{\phi^{[L]}_{\alpha}(\ell), \phi^{[R]}_{\alpha}(\ell)\}_{\alpha}$
respectively reducing the original $\phi_{\alpha}(\ell)$ to the left half and right half supports:
\begin{equation} \label{eq:planesplit}
    \phi^{[L]}_{\alpha}(\ell) = \sqrt{2}\; \Theta(L/2 - \ell) \;\phi_{\alpha}(\ell) \quad \mbox{and} \quad
    \phi^{[R]}_{\alpha}(\ell) = \sqrt{2}\; \Theta(\ell - L/2) \;\phi_{\alpha}(\ell), \vphantom{\int}
\end{equation}
with $\Theta$ being the Heaviside step function.
Even though in a general case a new set of wavefunctions generated via~\eqref{eq:planesplit}
would no longer be orthonormal, this specific choice of original orbitals $\phi_{\alpha}(\ell)$ preserves orthonormality:
\begin{equation} \label{eq:planesplit2}
  \begin{aligned}
    \sum_{\ell} \phi^{[L]}_{\alpha}(\ell) {\phi^{\star}}^{[R]}_{\beta}(\ell) &= 0 \qquad \mbox{and}\\
    \sum_{\ell} \phi^{[L]}_{\alpha}(\ell) {\phi^{\star}}^{[L]}_{\beta}(\ell) &=
    \sum_{\ell} \phi^{[R]}_{\alpha}(\ell) {\phi^{\star}}^{[R]}_{\beta}(\ell) = \delta_{\alpha, \beta}.
  \end{aligned}
\end{equation}
Let us write Fermi operators corresponding to this new set, satisfying the anticommutation rules
$ \{ \tilde{c}_{\alpha, L}, \tilde{c}_{\beta, L}^{\dagger} \} = 
\{ \tilde{c}_{\alpha, R}, \tilde{c}_{\beta, R}^{\dagger} \} = \delta_{\alpha, \beta}$,
all the other anticommutators being null.
It is clear that the original $\tilde{c}$ decompose in the new ones as
$\tilde{c}_{\alpha} = 2^{-1/2} (\tilde{c}_{\alpha,L} +  \tilde{c}_{\alpha,R})$,
thus allowing us to write the whole Slater determinant state as:
\begin{equation}
  |\Sigma\rangle = \frac{1}{2^{N/2}} \left( \tilde{c}^{\dagger}_{1,L} +  \tilde{c}^{\dagger}_{1,R} \right)
  \ldots \left( \tilde{c}^{\dagger}_{N,L} +  \tilde{c}^{\dagger}_{N,R} \right) |\Omega\rangle,
\end{equation}
By partially tracing over degrees of freedom related to one half of the system, say the right one,
we achieve the reduced density matrix of the left half
$\rho_{L}^{\Sigma} = \trace_{R}\left[ |\Sigma\rangle\langle\Sigma|\right]$.
With this goal in mind, just set $|\Sigma'\rangle = \tilde{c}_{1} |\Sigma\rangle$ and consider:
\begin{equation} \label{eq:reducedreduction}
  \begin{aligned}
    \rho^{\Sigma}_L &= \frac{1}{2}\; \trace_{R} \left[ ( \tilde{c}^{\dagger}_{1,L} +  \tilde{c}^{\dagger}_{1,R} )
      \;|\Sigma'\rangle \langle \Sigma'|\; ( \tilde{c}_{1,L} +  \tilde{c}_{1,R} ) \right] \\
    &= \frac{1}{2} \left( \tilde{c}^{\dagger}_{1,L} \trace_{R} \left[ |\Sigma'\rangle \langle \Sigma'|\right] \tilde{c}_{1,L}
    + \trace_{R} \left[ |\Sigma'\rangle \langle \Sigma'| \right] \right)
    = \frac{1}{2} \left( \tilde{c}^{\dagger}_{1,L} \:\rho^{\Sigma'}_L \:\tilde{c}_{1,L} + \rho^{\Sigma'}_L \right) 
  \end{aligned}
\end{equation}
where we exploited the cyclicity of the trace over right support operators $\tilde{c}_{\alpha,R}$, and the fact
that $\tilde{c}_{1,R} |\Sigma'\rangle = 0$.
Also it turns that $\rho^{\Sigma'}_L$ and $(\tilde{c}^{\dagger}_{1,L} \,\rho^{\Sigma'}_L\, \tilde{c}_{1,L})$ 
have orthogonal supports and the same spectrum\cite{NOTA2}.
Therefore we have $\eS_{\text{VN}}(\rho^{\Sigma}_L) = \eS_{\text{VN}}(\rho^{\Sigma'}_L) + \log 2$
[see pag.513 of Ref.~\refcite{NielsenChuang}, Theorem 11.8 (4)].

Finally, we repeat the argument~\eqref{eq:reducedreduction} on $|\Sigma'\rangle$, and proceed by induction. 
In conclusion, we can claim that $\rho^{\Sigma}_L$ is (isometrically equivalent to) $2^{-N} \Id_{2^{N} \times 2^{N}}$,
the maximally mixed state on a $2^{N}$ dimensioned space, having a von Neumann entropy of $N \log 2$.
This concludes the proof.

The previous result allows a clear and sensible interpretation:
fermions occupying the various orbitals must be mutually uncorrelated due
to the Slater determinant state nature, so the available entanglement
is given by the self-correlation of every orbital, separately accounted.
Rephrasing, there is no many-body entanglement in $|\Sigma\rangle$, but
one-body entanglement of each particle can still be present.
A single delocalized fermion can contribute to the total amount with
\emph{at most} the entanglement of a unit (i.e. the amount of entanglement shared by a spin singlet),
so $N$ units is naturally the overall maximum.

\section{Other exchange statistics}  \label{Sec:statistics}

It is important to remark that the $\sigma^{z}$ matrix in Eqs.~\eqref{eq:FermiMPOrecipe} 
is the only responsible for establishing the correct anticommutation relations of Fermi statistics.
That said, it is arguably easy to adjust~\eqref{eq:FermiMPOrecipe} in order to adapt
the representation to non-fermionic statistics\cite{KitaevAnyon}, 
as long as the particles involved are hard-core.
When a single level cannot be occupied by more than one particle (guaranteed by the Pauli
exclusion principle for fermions, must be inquired through other means otherwise)
then it is again possible to map the physical model into a 1D qubit lattice: 
the JWT of Eq.~\eqref{eq:Wigner1} actually becomes
\begin{eqnarray} \label{eq:Wigneralt}
  \ket{\Omega} & \longrightarrow & | 00 \ldots 0 \rangle \\
    c_{\ell}    & \longrightarrow & {\cal W}_{1} \otimes \ldots \otimes {\cal W}_{\ell-1} \otimes \sigma^{-}_{\ell}
  \otimes \Id_{\ell} \otimes \ldots \otimes \Id_L, 
  \nonumber
\end{eqnarray}
where ${\cal W}$ is the particle-exchange matrix which encodes the statistics. The $2 \times 2$ matrix
${\cal W}$ is equal to the identity $\Id$ for bosons, to the Pauli matrix $\sigma^{z}$ for fermions.
For abelian anyons which acquire a fractional phase $\varphi$ under exchange, ${\cal W}$ is still 
a diagonal matrix\cite{Corbozanyon}, precisely it becomes the phase gate ${\cal W} = \exp(i \varphi \sigma^{z})$.
Finally ${\cal W}$ can assume any shape when we consider particles whose exchange leads 
to the application of a non-trivial operator, that is non-abelian anyons\cite{KitaevAnyon}.

The generalization of the MPO representation~\eqref{eq:FermiMPOrecipe} of the creation 
operators is quite straightforward for bosonic and {\it abelian} anyonic scenarios. 
One needs only to change the matrix ${B^{[\ell]}}_{1}^{1}$ as
\begin{equation}
  {B^{[\ell]}}_{1}^{1} = {\cal W}
\end{equation}
and keep the rest. Since now ${\cal W}$ is the phase gate (with trivial phase $\varphi = 0$
if we are considering bosons), it is still diagonal in the eigenbasis of $\sigma^{z}$, thus meaning 
that the matrix product contraction with the same boundaries $( \beta_0 | = (1|$ and $| \beta_{L} ) = |0)$ 
is reduced like in Eq.~\eqref{eq:firstreux}, i.e.
\begin{equation} \label{eq:secondreux}
  \sum_{\ell} \phi_{\alpha}(\ell) \left[ {\cal W}_1 \otimes \ldots \otimes {\cal W}_{\ell-1}
    \otimes \sigma_{\ell}^{-} \otimes \Id_{\ell+1} \otimes \Id_{L} \right]
  \\ \longrightarrow
  \sum_{\ell} \phi_{\alpha}(\ell)\, a_{\ell}^{\dagger} = \,\tilde{a}_{\alpha}^{\dagger},
\end{equation}
with $a_{\ell}^{\dagger}$ ($a_\ell$) being the abelian anyonic/bosonic construction (destruction) 
operator on site $\ell$, while $\tilde{a}_{\alpha}^{\dagger}$ ($\tilde{a}_{\alpha}$) 
the construction (destruction) operator for the particle orbital $\alpha$.
We can then write the MPS representation $A$ for the hard-core many-body state \emph{free} 
of correlations (generalization of the Slater Determinant for an arbitrary abelian exchange 
statistics ${\cal W} = e^{i \varphi \sigma^{z}}$) as follows
\begin{eqnarray} \label{eq:Slamprecipe3}
    A^{[\ell]}_0 & = & \Id_{2^N \times 2^N} \\
    A^{[\ell]}_1 & = & \sum_{\alpha = 1}^{N} \phi_{\alpha}(\ell)
    \left( {\cal W}^{\otimes {N-\alpha}} \otimes \sigma^{-} \otimes {\Id}_{2 \times 2}^{\otimes {\alpha -1}}
    \right). \nonumber
\end{eqnarray}

Unfortunately, the non-abelian anyonic case is more difficult to be addressed:
since the general form of a particle-exchange matrix ${\cal W}$ does not commute with $\sigma^{z}$
the MPO construction cannot eliminate rigorously multi-particle creation events, 
therefore the previous generalization is not exploitable.
In this case, it is not clear to us whether a larger refinement parameter $D$ is required.

\section{From Slater determinants to configuration interaction picture}  \label{Sec:CI}

More accurate results can be achieved by adopting the Configuration Interaction (CI) picture. 
In this description, Hartree-Fock solutions are adopted as a canonical vector basis of orbitals 
for further calculations, embedding entanglement between particles\cite{CI1999}.
According to such approach, one is interested to express quantum correlations by superposing
few to several Slater determinant states, which typically \emph{share} most of the HF orbitals while differing
for a limited amount of those, in order to keep the calculation manageable for practical purposes.

Having this scheme in mind, we would like to extend our previous Slater MPS (via MPO stack) representation to embed
also CI states, where different orbital excitations are coherently overlapped.
The ultimate ingredient of this perspective would be writing exact Matrix Product representations for every operator
generated by the Fermi ones $\tilde{c}^{\dagger}_{\alpha}$ through sums and multiplications.
Unfortunately, providing a general, elegant and high-compatibility formulation is no easy task,
since the matrix product representations are likely to be operator-dependant and definitely not unique.
Here we will consider the simplest nontrivial case, and speculate on various proposals.

For instance, consider the expression
\begin{equation}
  \Theta_{2+2} = \alpha \;\tilde{c}^{\dagger}_1 \,\tilde{c}^{\dagger}_2 +
  \beta \;\tilde{c}^{\dagger}_3 \,\tilde{c}^{\dagger}_4.
\end{equation}
We are now going to describe $\Theta_{2+2}$ as a MPO.
Depending on whether we focus on the adaptability of the description or the economy on the refinement parameter
dimension we can end up with different representations.

In the first case one exploits the standard technique to sum coherently Matrix
Product objects, and is strongly based on the MPO structure of Fermi operators in 
Eqs.~\eqref{eq:FermiMPOrecipe}. It is highly suitable for further generalizations,
but at the cost of a sub-optimal bond-link dimension.
Let us now adopt $D = 8$ and consider
\begin{equation} \label{eq:blockdj}
  {B^{[\ell,2+2]}}_{j}^{i} = \sum_k \left( \begin{array}{c|c}
    {B^{[\ell,1]}}_{k}^{i} \otimes {B^{[\ell,2]}}_{j}^{k} & 0 \\ \hline
    0 & {B^{[\ell,3]}}_{k}^{i} \otimes {B^{[\ell,4]}}_{j}^{k}
  \end{array} \right),
\end{equation}
where the $B^{[\ell,\alpha]}$ tensors are those defined in~\eqref{eq:FermiMPOrecipe} for $\tilde{c}^{\dagger}_{\alpha}$.
The basic idea behind this construction is to use a correlation space which is the
Cartesian sum of the two original correlation spaces, and a matrix product object which is the block
diagonal composition. Similarly we define the correlation boundary vectors, 
which contain information on $\alpha$ and $\beta$:
\begin{equation}
  | b_L ) =
  \mbox{\scriptsize{$ \left( \begin{array}{c} \alpha \\ 0 \\ 0 \\ 0 \\ \beta \\ 0 \\ 0 \\ 0 \end{array} \right) $}}
  = \alpha
  \mbox{\footnotesize{$ \left( \begin{array}{c} 1 \\ 0 \\ 0 \\ 0 \end{array} \right) $}}
  \oplus \beta
  \mbox{\footnotesize{$ \left( \begin{array}{c} 1 \\ 0 \\ 0 \\ 0 \end{array} \right) $}} 
  = \mbox{\footnotesize{$ \left( \begin{array}{c} \alpha \\ \beta \end{array} \right) $}}\otimes
  \mbox{\footnotesize{$ \left( \begin{array}{c} 1 \\ 0 \\ 0 \\ 0 \end{array} \right) $}},
\end{equation}
where we used distributivity of the tensor product $\otimes$ with respect to the Cartesian sum $\oplus$.
Similarly, $(b_0| = ($\mbox{\tiny{1 1}}$) \otimes ($\mbox{\tiny{0 0 0 1}}).

On the contrary one might also like to keep the lowest possible value of the 
refinement parameter, $D = 6$. In this case one shall take
\[
 {B^{[\ell,2+2]}}_{0}^{0} = \Id_{6 \times 6} \, ,\qquad {B^{[\ell,2+2]}}_{0}^{1} = 0 \, ,
\]
\[
 {B^{[\ell,2+2]}}_{1}^{0} \!= \! \left(
 \begin{array}{cccccc}
 0 & 0 & 0 & 0 & 0 & 0 \\
 - \alpha \phi_1(\ell) & 0 & 0 & 0 & 0 & 0 \\
 - \beta \phi_3(\ell) & 0 & 0 & 0 & 0 & 0 \\
 \beta \phi_4(\ell) & 0 & 0 & 0 & 0 & 0 \\
 \alpha \phi_2(\ell) & 0 & 0 & 0 & 0 & 0 \\
 0 & \phi_2(\ell) & \phi_4(\ell) &
 \phi_3(\ell) & \phi_1(\ell) & 0
 \end{array}
 \right)
\]
\begin{equation}  \label{eq:cheapyMPO6}
  \mbox{and} \qquad
       {B^{[\ell,2+2]}}_{1}^{1} = \left( \begin{array}{cccccc}
         1 & 0 & 0 & 0 & 0 & 0 \\
         0 & -1 & 0 & 0 & 0 & 0 \\
         0 & 0 & -1 & 0 & 0 & 0 \\
         0 & 0 & 0 & -1 & 0 & 0 \\
         0 & 0 & 0 & 0 & -1 & 0 \\
         0 & 0 & 0 & 0 & 0 & 1
       \end{array} \right),
\end{equation}
while boundaries are left as before $|b_L) = |0)$ and $(b_0| = \left( \mbox{\tiny{0 \ldots 0 1}} \right) = (5|$.
By multiplying the $B^{[\ell,2+2]}$ matrices it is easy to see that we are reproducing the correct action of
the operator, i.e.
\begin{equation} \label{eq:sommy1}
  \sum_{\ell_1 < \ell_2} \left\{ \alpha \left( \vphantom{\sum} \phi_1(\ell_1) \phi_2(\ell_2) 
  - \phi_2(\ell_2) \phi_1(\ell_1) \right) + 
  \beta \left(  \vphantom{\sum} \phi_3(\ell_1) \phi_4(\ell_2) - \phi_4(\ell_2) \phi_3(\ell_1) \right) \right\}
  \, c^{\dagger}_{\ell_1} \, c^{\dagger}_{\ell_2}.
\end{equation}
One can ask whether this Matrix Product representation is optimal in terms of correlation bond-link dimension.
A state of the form $\Theta_{2+2} |\Omega\rangle$ manifests an entanglement entropy
\emph{at most} equal to $5/2$. This implies that a faithful MPS description would
require a $D \geq \sqrt{32}$, so that $D = 6$ is the smallest allowed integer, and thus arguably optimal.

The present proposal presents various options for generalization, although finding the analytical MPO expression
for a generic operator which is cheapest in terms of $D$ is definitely non-trivial.

\section{Conclusions}  \label{Sec:concl}

In this paper we studied matrix product representations for uncorrelated many-body states,
expressed in an arbitrary configuration space upon which the orbitals might be highly delocalized.
To achieve this goal, we provided the exact representation of a (delocalized orbital) creation operator
as an MPO, which we checked to be optimal in terms of the refinement parameter. Stacking these
MPO leads to the description of the uncorrelated many-body states quite automatically.
The representation we introduced is tailored upon fermionic exchange statistics, but it generalizes easily to
particles obeying any abelian exchange statistic. Also, it leads to a compact and simple tensor network
representation of the many body transformation corresponding to one-body basis change.
Finally, we discussed some possible extension of the previous description to embed also states carrying
a controlled amount of many-body correlation, such as CI states. We provided some proposals for the
simplest cases and argued that there is no unique, highly-compatible and optimal solution.

We conclude by arguing that, while the tensor network representation proposal we tailored in this paper
was based on a one-dimensional qubit lattice, we are positive that similar construction
can be adapted to other lattice geometries, as long as some peculiar topological features are preserved.
In particular, we are thinking about finitely correlated states on Cayley trees\cite{Cayley}, 
sharing with 1D OBC systems the fact that no closed loops in the geometry are present.

\section*{Acknowledgments}

PS acknowledges support from EU through AQUTE and PICC.
DR acknowledges support from EU through SOLID.
GES acknowledges support by the Italian CNR, through ESF Eurocore/FANAS/AFRI,
by the Italian Ministry of University and Research, through PRIN/COFIN 20087NX9Y7,
by the SNSF, through SINERGIA Project CRSII2 136287/1, and by the EU-Japan Project LEMSUPER.
VG acknowledges support by MIUR through the FIRB-IDEAS project No. RBID08B3FM.


\begin{thebibliography}{1}
\bibitem{NOTA1} 
  Our results naturally extend to fermions with spins, by replacing the bare orbital sets with spin-orbitals.
\bibitem{NOTANEW1}
  In writing Eq.~\eqref{eq:MPSampli} we are implicitly selecting open boundary conditions, 
  encoded in the presence of correlation space boundaries $|b_L)$ and $(b_0|$. 
  The generalization to the case of periodic boundary conditions is trivial, but not particularly convenient.
\bibitem{NOTA2}
  This follows from the fact that $\rho^{\Sigma'}_L$ differs from 
  $(\tilde{c}^{\dagger}_{1,L} \,\rho^{\Sigma'}_L\, \tilde{c}_{1,L})$ only in the population 
  of the orbital $\phi_1^{[L]}$, being it empty in the first case and occupied in the second case.
\end{thebibliography}
\end{document}